\documentclass{webofc}

\usepackage[varg]{txfonts}   
\usepackage{hyperref}
\usepackage{url}
\usepackage[T1]{fontenc}
\usepackage[utf8]{inputenc}
\hypersetup{colorlinks=true,citecolor=blue,urlcolor=blue,linkcolor=blue}
\begin{document}
\title{Challenges for Monte Carlo generators}
%
%

\author{\firstname{J\"urgen} \lastname{Reuter}\inst{1}\fnsep\thanks{\email{juergen.reuter@desy.de}}
}

\institute{Deutsches Elektronen-Synchrotron DESY, Notkestr 85, 22607
  Hamburg, Germany}

\abstract{This contribution lists challenges of Monte Carlo event
  generators for future lepton, especially linear colliders. A lot of
  the recent development benefits from the achievements at the Large
  Hadron Collider (LHC), but several aspects are unique for lepton
  colliders like beam simulation, polarization, electroweak higher
  order corrections and resummed QED corrections. We will describe the
  status of multi-purpose event generators and specialized codes and
  outline the challenges for these tools until such a collider starts
  data taking.}
\maketitle
\section{Introduction}
\label{intro}
Monte Carlo (MC) event generators for lepton colliders (both
electron-positron and muon colliders, but also photon colliders) are
not conceptually different than for hadron colliders. They simulate a
hard scattering process in an order of perturbation theory as high as
possible, however, they are dominated by QED and electroweak instead
of QCD corrections, and hence theoretical uncertainties due to large
scale variations are (almost) not present. Parton distribution
function (PDF) uncertainties are either not present or can be
perturbatively estimated. QCD parton showers and hadronization proceed
in the same as at LHC, however without multiple interactions and
underlying event (though there are overlays from the bunch
structures). In the following sections, we will discuss first discuss
the simulation of beam spectra in Sec.~\ref{sec:beamspectra}, then
hard matrix elements and the treatment of initial-state radiation in
Sec.~\ref{sec:hard_initial}, parton showers, matching and
hadronization are briefly reviewed in Sec.~\ref{sec:shower_hadro}. The
next Sec.~\ref{sec:special} deals with specific processes that
need a specialized treatment and sometimes specialized tools; in
Sec.~\ref{sec:bsm} we quickly discuss simulation of BSM processes,
while in Sec.~\ref{sec:perform} we review phase space sampling,
performance issues and parallelization, before we conclude.

This review is oriented along the lines of the ECFA
Higgs-Top-Electroweak Factory studies, carried out 2022-2024, and the
developments of MC generators there: according to this study, the
focus here is on the the three multi-purpose multi-leg generators
Magraph5\_aMC@NLO (MG5)~\cite{Alwall:2014hca},
Sherpa~\cite{Gleisberg:2003xi,Sherpa:2019gpd} and 
Whizard~\cite{Kilian:2007gr,Moretti:2001zz}, the two 
multi-purpose shower/hadronization programs
Herwig~\cite{Bahr:2008pv,Bellm:2015jjp}
and
Pythia6/8~\cite{Sjostrand:2006za,Sjostrand:2014zea,Bierlich:2022pfr},
and the specialized tools BabaYAGA~\cite{CarloniCalame:2003yt},
BHLumi/BHWide~\cite{Jadach:1996is,Jadach:1995nk}, and 
KKMC~\cite{Jadach:2022mbe}. An excellent overview is given in the
input document for the US Snowmass Community
Study~\cite{Campbell:2022qmc}. Many interesting details on the
simulation of the complete SM processes for ILC can be found
in~\cite{Berggren:2021sju}. Details on the simulation can also be
found in the conceptual or technical design reports of the
electron-positron and muon
colliders~\cite{CEPCStudyGroup:2023quu,ILC:2013jhg,Behnke:2013lya,FCC:2018evy,Aicheler:2012bya,Linssen:2012hp,Accettura:2023ked}. 


\section{Beam spectrum simulations}
\label{sec:beamspectra}

High-luminosity lepton colliders exhibit beamstrahlung, radiation from
the bunch of electrons (or positrons) in the electromagnetic field of
the approaching other colliding, highly collimated bunch. For
synchrotrons, this effect is much less severe for the colliding
bunches, and a Gaussian-shaped beam energy spread is mostly
sufficient. Gaussian beam spreads are available in most Monte Carlos
like e.g. KKMC, Pythia8, and Whizard. To cover a bit more features is
possible by approximating the beam spectra by smeared delta-peak like
structures with power-law tails, making the assumptions that the beam
spectrum of the two lepton beams factorizes into two multiplicative
components. This approach has been codified in the tool
CIRCE1~\cite{Ohl:1996fi} that was used for the simulations of the
TESLA linear collider project. This approach is available in KKMC,
MG5~\cite{Frixione:2021zdp}, and Whizard, with slight variations in
the parameterization. However, it has been shown that for ILC and
drive-beam accelerators like CLIC, plasma accelerators like HALHF as
well as photon colliders, such a parameterization is
insufficient, as both the peak structure has a very complicated shape
and also the tails do not exhibit simply a power-like behavior. Here,
an MC generator based on a two-dimensional histogrammed fit to
accelerator spectra simulated by tools like Guinea-Pig or Cain. Due to
their limited statistics, a special smoothening has to be applied to
avoid artifacts. Also the peaked parts of the spectra close to the
full energy have to be special-cased in order to avoid unphysical beam
energy spreads. This is realized in the CIRCE2 algorithm in the
Whizard generator, where spectra for CEPC, CLIC, and ILC are
available, while the FCC-ee spectra will be made available soon. Note
that besides the $e^+e^-$ components of the spectra, also the
$e^+\gamma$, $\gamma e^-$ and $\gamma\gamma$ components are very
important, especially for linear colliders ($\gamma\gamma$-induced
backgrounds).

Further developments in the future comprise the first realistic
spectra for plasma-driven accelerators like HALHF or energy-recovery
LINACs (ERLs), the 3-dimensional structure of beam spectra, especially
their dependence along the beam axis (e.g. using so-called copulas),
and the use of advanced machine learning algorithms for the fitting of
beam spectra from accelerator simulation tools.


\section{Hard matrix elements and resummation of initial-state radiation}
\label{sec:hard_initial}

The structure of hard scattering amplitudes from the (multi-leg)
matrix element generators are very similar to the corresponding LHC
amplitudes. These are available at leading (LO) and next-to-leading
order (NLO) fully automatized, and rely on recursive algorithms,
cf.~e.g.~\cite{Ohl:2023bvv}. A comparison for multi-leg EW production
processes at the LHC can be found e.g. in~\cite{Ballestrero:2018anz}.
Obviously, compared to the LHC NLO QCD corrections are of minor
importance as they do not lead to order-of-magnitude changes due to
large scale variations. NLO electroweak (EW) corrections dominate due
to phyics processes dominated by EW resonance production. Examples for
NLO QCD corrections to basically all important processes can be found
in~\cite{Alwall:2014hca,Rothe:2021sml,Bredt:2022nkq,Stienemeier:2022wmy},
while examples for NLO EW corrections can be found
in~\cite{Kilian:2006cj,Robens:2008sa,Bredt:2022dmm}. This has been
made possible by the vast amounts of progress at the level of one-loop
(matrix-element) providers like GoSam~\cite{GoSam:2014iqq},
MadLoop~\cite{Alwall:2014hca},
Openloops~\cite{Cascioli:2011va,Buccioni:2019sur} and
Recola~\cite{Actis:2016mpe}, using the BLHA 
interface~\cite{Binoth:2010xt,Alioli:2013nda}. Automated subtraction
schemes like Catani-Seymour (CS)~\cite{Catani:1996vz} and
Frixions-Kunszt-Signer (FKS)~\cite{Frixione:1995ms,Frixione:1997np} take care
of the numerical cancellation of infrared singularities in four
spacetime dimensions where the MC integrations are performed. Generic
subtraction schemes for at least QCD at the NNLO level are under
active development.

The second major bottleneck to go to NNLO besides the automated NNLO
subtraction, is the evaluation of two-loop master integrals which for
EW processes at e.g. ILC or FCC-ee contains several different mass
scales, and the frontier lies at 4- and 5-point functions with two
different mass scales. First baby steps towards tools for simple
processes like $e^+e^- \to ZH$ (on-shell) have started, but it will
take a still many years until that is nearly as mature as for NLO.
There is another bottleneck, which is unavoidable for fixed-order NLO
or NNLO calculations, namely the appearance of negative-weight events
which makes sending them to a (fast) detector simulation
computationally extremely expensive. This is, of course, well-known
from NLO and NNLO QCD MC simulations at the LHC, but plagues NLO QCD
and NLO EW calculations at lepton colliders in the same way. This has
been shown to apply especially to the top threshold simulation which
will be described below in Sec.~\ref{sec:special}.

To achieve permil-level precision on total cross sections and
differential distributions, the resummation of soft and collinear
photon radiation in the initial state is mandatory. This has started
from the soft-collinear resummation to all
orders~\cite{Gribov:1972ri}, with the inclusion of hard-collinear
photons up to second~\cite{Kuraev:1985hb} and third order~\cite{Skrzypek:1990qs}. In
general, there are the two options to use either collinear factorization
resumming collinear logarithms (at LL~\cite{Cacciari:1992pz} or at
NLL~\cite{Frixione:2019lga,Bertone:2019hks,Bertone:2022ktl}) or eikonal 
factorization resumming soft logarithms (also called YFS
exponentiation~\cite{Yennie:1961ad,Jadach:2000ir,Krauss:2022ajk}).
In an ideal world, both types of logarithms have to be combined to
achieve the highest possible precision. While there is an algorithmic
procedure to improve YFS exponentation (which also can generate
exclusive, though mostly soft, photons) by hard-collinear corrections,
collinear factorization seems to be better suited for combination with
higher fixed-order calculations of the hard process. Very likely, each
of the methods has its own benefits and which is better suited for
what processes needs to be studied in the future. For a nice
summaries on these topics, cf. also~\cite{Heinemeyer:2021rgq,Frixione:2022ofv}.

It should not be forgotten that the simulation of polarized processes
(both in the initial state, but also in the final state for polarized
event samples) is a must for the physics program of linear lepton
colliders. The most general formalism here is based on spin-density
matrices which allow arbitrary polarizations in the initial state,
e.g. both longitudinal and transversal polarization of lepton beams.


\section{Parton showers, matching, and hadronization}
\label{sec:shower_hadro}

Parton showers resum large logarithms and provide exclusive multi-jet
events; again, parton showers have profited tremendously from two
decades of developments for LHC. This has driven the development of
(final-state) showers that are accurate at next-to-leading logarithmic
(NLL) order, and hence include effects from color and spin
correlations beyond the quasi-classical approximation of independent
emissions. Examples of these
are~\cite{Dasgupta:2020fwr,Herren:2022jej,Nagy:2020rmk,Forshaw:2020wrq}. These
showers are available for jet distributions at linear colliders and
have been e.g. applied to hadronic Higgs events at future lepton
colliders~\cite{Knobbe:2023njd} or reapplied to LEP
data~\cite{Kilian:2011ka}. For lepton colliders, QED showers, and for
high-energy linear lepton colliders like ILC and CLIC also electroweak
showers do become important. These are mostly realized as interleaved
showers where the QED/EW suppression of emissions from quarks by a
factor of $\alpha/\alpha_s ~ 1/15$ is dealt with by a combination of
rejection and reweighting algorithms. For EW showers, the question of
EW fragmentation of inclusive W/Z/H radiation into especially hadronic
jets with jet masses around the EW resonances will become important.
Techniques and algorithms to match fixed-order calculations to
exclusive all-order showers have been developed for hadron colliders
for LO (matrix element corrections in parton showers) to NLO -- MC@NLO,
POWHEG~\cite{Frixione:2007vw}, CKKW(-L) -- to NNLO (MiNNLOPS, UnLLOPS
etc.) and even NNNLO, while consistently merging samples of exclusive
jet multiplicities into inclusive multi-jet merged samples have been
developed for the LHC. These techniques can be straightforwardly carried over to lepton
colliders like ILC and CLIC. While the corresponding matching of
photon emission (and lepton pairs) with (N)NLO EW corrections is in
principle straightforward, the complete automation is still work in
progress and will only be available in the future. A matching schemes
at LO for matrix element and shower photons can be found
in~\cite{Kalinowski:2020lhp}. 

Hadronization is still based mostly on phenomenological models, either
on Lund string fragmentation or the cluster hadronization
paradigm. Machine-learning methods aim at simulating hadronization
from trained real data event samples, which might result in realistic
hadronic event generation, but, however, does not enhance our
knowledge about the underlying QCD physics. There are speculations
that samples of up to 200 ab${}^{-1}$ of very clean jet samples from
e.g. FCC-ee might necessitate the development of new formalisms for
our understanding of hadronization in order to understand the data.


\section{Special processes and dedicated tools}
\label{sec:special}

There are several processes that exhibit special properties,
(i) two-fermion production $e^+e^- \to f\bar{f}$, (ii) Bhabha
scattering $e^+e^- \to e^+e^-$, two-photon production $e^+e^- \to
\gamma\gamma$, (iii) photoproduction of low-$p_T$ hadrons, (iv) the
$WW$ threshold, (v) the top threshold. The first two are important for
high-precision luminometry, where linear colliders favor Bhabha
scattering as the lumical and beam calorimeter allow to measure much
more inclusive to very small angles. Matching the projected
experimental precision of $10^{-4}-10^{-5}$ theoretically is very
challenging and needs a dedicated implementation of complete EW
two-loop corrections and leading three-loop corrections together with
resummation of all available soft and collinear logarithms. This has
been achieved in dedicated tools like BabaYaga, BHLumi and BHWide. The
photoproduction of low-$p_T$ hadrons is one of the largest background
processes especially in the central region and for high-energy
versions like CLIC. Its simulation is based on fits to total cross
section measurements from the CrystalBall experiment, while the
hadronization for low effective center-of-mass energies is dominated
by pion and kaon production. A special implementation exists e.g. in
Whizard. For the $WW$ threshold a very precise determination of the
$W$ mass necessitates the resummation of QED logarithms which can be
added to a fixed-order calculation. More complicated is the top
threshold where inclusive results are available at NNNLO NRQCD that
will allow an extraction of the top mass from the threshold scan at
the level of 30-50 MeV uncertainty. For the study of experimental
acceptances and systematic uncertainties, this has to be included in
an exclusive MC, which can be achieved at level of NLL matched to NLO
NRQCD~\cite{Bach:2017ggt,ChokoufeNejad:2016qux}. Carrying this over to
NNLL matched to NNLO would be desirable, but will be a major
undertaking. 


\section{Simulation of beyond the Standard Model physics}
\label{sec:bsm}

Also the simulation of BSM processes builds upon 15 years of
development for LHC BSM simulations, relying on Lagrangian-level tools
like SARAH~\cite{Staub:2013tta}, LANHep~\cite{Semenov:2008jy} and
FeynRules~\cite{Alloul:2013bka}, that are connected to MC simulations 
via dedicated interfaces like e.g. in~\cite{Christensen:2010wz}. These
tailor-made interfaces that have to be engineered for every pair of
these tools with each MC generator. They have become redundant through
the introduction of a python-syntax based meta-layer within the
Universal Feynman Output (UFO)
interface~\cite{Degrande:2011ua,Darme:2023jdn}. For lepton colliders
like the ILC and CLIC especially models with light EW particles like
in extended Higgs sectors, axions and axion-like particles (ALPs),
heavy neutral leptons etc. are very important. In addition, deviations
of the SM are parameterized in terms of effective field theories like
SMEFT or HEFT, which are available as well for such simulations. There
are many attempts to carry these simulations over to NLO; for NLO QCD
this is possible for most of these models as mandated to achieve a
decent precision for LHC signal simulations; for NLO EW calculations
this is much harder to achieve generically as proper renormalization
schemes have to be available which can very likely never be fully
automatized. Also colorful exotics are available, in direct production
processes for the LHC or in terms of EFT operators for ILC/CLIC as
they appear in Higgs portal and dark sector models. Details on their
implementation based on the color-flow formalism can be found
e.g. in~\cite{Ohl:2024fpq,Kilian:2012pz}.


\section{Phase space, parallelization and performance}
\label{sec:perform}

Physics simulations are computationally costly, especially at NLO and
NNLL. As lepton colliders like ILC and CLIC (or CEPC and FCC-ee)
exhibit processes that are dominated by EW resonances, simulation
usually is classified by the number of the fermions in the final
state: $e^+e^-/e\gamma/\gamma\gamma \to 2f, 3f, 4f, 5f, 6f, 7f, 8f,
\ldots$. These processes can become very costly with tens of thousands
of phase space channels already at LO, but much more so at NLO. This
already shows that phase spaces become much more complicated at lepton
colliders with many intertwined electroweak production channels
connected by (EW) gauge invariance compared to phase spaces dominated
by soft-collinear QCD radiation phenomena at LHC. Such processes need
to be sampled in parallel and event generation needs to be treated
parallelized as well. The latter is always trivially achievable as
bundles of events can be generated on different nodes of large
computer farms. There are straightforward algorithms for parallelized
matrix element evaluation, e.g. parallelizing over the helicities of
external particles ($2^n$ for $n$ external fermions) or the
parallelization of color flows over different threads in
hyper-threading. Phase space adaptation is much more complicated: not
only need random number chains be independent on different nodes (as
for event generation) but also correlations between different parts of
the phase-space adaptation need to be taken into account. One example
which shows speed-ups of 30-50 is the parallelization of multi-channel
adaptive VEGAS-type phase space sampling using the message-imaging
protocal (MPI)~\cite{Brass:2018xbv}. Also, the usage of GPUs for
matrix-element evaluation as well as phase-space sampling shows
interesting speed-ups, and there are attempts in all three
multi-purpose generators MG5, Sherpa and
Whizard~\cite{Hagiwara:2013oka,Valassi:2021ljk,Bothmann:2021nch}. Further
interesting developments is the phase-space sampling using
machine-learning techniques like invertible neural networks,
normalizing flows or autoencoders, which is left out here for the sake
of space.

\section{Summary and outlook}
\label{sec:summary}

MC generators for lepton colliders (electron-positron and muon
colliders) build upon two decades of development for LHC: these have
culminated in the automation of NLO QCD and NLO EW calculation for
arbitrary multi-leg processes. While this is almost trivially carried
over for NLO QCD, for NLO EW it is tremendously complicated due to the
combination with the numerically demanding NLL QED PDFs. Future
studies will show which processes and energies are in the regime of
collinear radiation or of soft radiation which is simulated in the YFS
formalism. Another future task is a universally applicable matching
formalism between fixed-order and resummed calculations with exclusive
radiation simulations and QED showers. This is mandated to achieve
permil precision for both inclusive and exclusive predictions. Very
high-energy lepton (especially muon) colliders might call for a full
EW/SM collinear factorization using EW PDFs and corresponding EW
fragmentation functions, which will be a very interesting field of
research in the future.
The inclusion of beam spectra are important for the full simulation of
physics processes for the future experiments at lepton colliders. It
has been shown that the most powerful is a two-dimensional
histogrammed fit to beam spectra, that allows to properly describe
synchrotrons like CEPC and FCC-ee, RF-based accelerators like ILC and
C3, drive-beam accelerators like CLIC, plasma accelerators like HALHF
and even photon colliders.
Parton showers have matured in the LHC era, being available at NLL
now, given access to non-trivial color and spin correlations, and will
likely be carried to full NNLL in the future. Hadronization models
might have to be rethought in an era with gigantic samples of very
clean hadronic data from the Z pole.
BSM simulations profit again from the LHC developments, and are
available for multi-purpose MCs for almost arbitrary BSM models using
Lagrangian-level tools connected to MCs via the UFO2 python layer. A
challenge will be EW NLO corrections in arbitrary BSM models, as they
always depend on the availability of appropriate renormalization
schemes.
There are several processes that need special treatment either because
they are needed at a much higher theoretical precision than the rest
or they are very delicate in terms of soft and collinear
divergencies. These processes are very often described by dedicated
tools, while sometimes multi-purpose generators contain specialized
treatments of such processes. Examples are processes for luminometry,
the $WW$ and top threshold and photon-induced production of low-$p_T$
hadrons.
MC simulations have been understood as one of the computational
bottlenecks of the data analysis at LHC, and this still remains true
at lepton colliders, especially at Tera-Z. Active areas of research
are phase-space sampling improvements based on either adaptive
multi-channel versions of VEGAS or machine-learning methods based on
e.g. normalizing flows. Another very active field is the usage of
heterogeneous computing where parts of the simulation remains on
traditional CPU farms and parts are off-loaded to GPUs.

\section*{Acknowledgments}

For this work we would thank the organizers for a fantastic conference
with stimulating atmosphere. The author acknowledges funding by the Deutsche
Forschungsgemeinschaft (DFG, German Research Foundation) under grant
396021762 - TRR 257, under grant 491245950 and under Germany's
Excellence Strategy-EXC 2121 ``Quantum Universe"-390833306, by
National Science Centre (NCN, Poland) under the OPUS research project
no. 2021/43/B/ST2/01778 and by EAJADE - Europe-America-Japan
Accelerator Development and Exchange Programme (101086276), and from
the International Center for Elementary Particle Physics (ICEPP), the 
University of Tokyo. 

\bibliography{mcgen_jrr}

\end{document}